\documentstyle[aps,preprint]{revtex}
\begin{document}
\tighten

\title{\bf The many faces of superradiance}
\author{Jacob D. Bekenstein\footnote[1]{ E--mail: bekenste@vms.huji.ac.il}
and Marcelo Schiffer\footnote[2]{Also at the
Department of Applied Mathematics and Statistics, University of
Campinas, Campinas SP, Brazil} \footnote[3]{E--mail:
schiffer@obelix.unicamp.br}}
\address{\it Racah Institute of Physics, Hebrew University of
Jerusalem\\ Givat
Ram, Jerusalem 91904, Israel}
\date{Received \today}
\maketitle
\begin{abstract}
Inertial motion superradiance, the emission of radiation by an initially
unexcited system moving inertially but superluminally through a medium, has
long been known. Rotational superradiance, the amplification of radiation by a
rotating rigid object, was recognized much later, principally in connection
with black hole radiances.  Here we review the principles of inertial motion
superradiance and prove thermodynamically that the Ginzburg--Frank condition
for superradiance coincides with the condition for superradiant amplification
of already existing radiation. Examples we cite include a new type of 
black hole superradiance.  We correct Zel'dovich's thermodynamic derivation of
the Zel'dovich--Misner condition for rotational superradiance by including the
radiant entropy in the bookkeeping .  We work out in full detail the
electrodynamics of the Zel'dovich rotating cylinder,  including a general
electrodynamic proof of the Zel'dovich--Misner condition, and explicit
calculations of the superradiant gain for both types of polarization. 
Contrary to Zel'dovich's pessimistic conclusion we conclude that, if the 
cylinder is surrounded by a dielectric jacket and the whole assembly is placed
inside a rotating cavity, the superradiance is measurable in the
laboratory.

\end{abstract}


\section{INTRODUCTION}  \label{intro}


A free structureless particle moving inertially in vacuum cannot absorb or
emit a photon.   This well known fact follows solely from Lorentz
invariance and four--momentum conservation.  But a free object endowed
with internal structure can, of course, absorb photons, and can also emit
them provided it is initially excited above its ground state (rest
mass $M$ larger than minimum possible value $M_{\rm gr}$).  Somewhat
surprisingly, when the object, which may be electrically neutral overall,
moves uniformly through a medium, emission may be allowed even when the
object starts off in its ground state ! Ginzburg and Frank's early
recognition of this possibility \cite{GinzFr} (Ginzburg \cite{Ginz} gives a
modern review) marks the beginning of our subject, which we term
superradiance.

The term superradiance, introduced by Dicke \cite{Dicke}, originally referred
to amplification of radiation due to coherence in the emitting medium.   Many
years later Zel'dovich \cite{Zeld1} pointed out that a cylinder
made of absorbing material and rotating about its axis with frequency $\Omega$
is capable of amplifying those modes of scalar or electromagnetic radiation
impinging on it which satisfy the condition
\begin{equation}
\omega - m \Omega < 0
\label{supercriterion}
\end{equation}
where $\omega$ is the waves' frequency and $m$ the azimuthal quantum number
with respect to the axis of rotation.  Zel'dovich realized that, when quantum
physics is allowed for, the rotating object should be able to emit
spontaneously in the regime (\ref{supercriterion}), and anticipated
that a rotating (Kerr) black hole should show both amplification and
spontaneous emission when condition (\ref{supercriterion}) is satisfied.  
Misner \cite{Misner} independently made a suggestion that the Kerr black hole
will amplify waves, and supported it with unpublished calculations.   The
corresponding spontaneous emission was first put into evidence
field--theoretically by Unruh
\cite{Unruh}.

Following Misner's observation one of us noted \cite{Bek_super} that in the
Kerr black hole case superradiant amplification is classically required  when
condition (\ref{supercriterion}) holds because that is the only way to fulfill
Hawking's classically rigorous horizon area theorem \cite{Hawking_area} (see
also Ref.~\cite{PressTeuk1}). From the same logic it followed \cite{Bek_super}
that superradiance of electrically charged waves by a charged black hole is
required whenever
\begin{equation} 
\omega -q \Phi/\hbar < 0
\label{chargedcriterion}
\end{equation}
where $q$ is the elementary charge of the field and $\Phi$ the
electrostatic potential of the black hole measured at the horizon.

Following the emergence of black hole thermodynamics
\cite{BHThermo} it became clear that black horizon area plays the
role of entropy for black holes.   This correspondence and the
cited argument for superradiance from black hole area immediately suggests
that the necessity of superradiance in ordinary objects is
solely a consequence of thermodynamics.   In fact, Zel'dovich \cite{Zeld2}
used a thermodynamic argument in his discussion to show that  superradiance of
the rotating cylinder must take place.  Following this idea we extend in this
paper the superradiance condition to a broad range of circumstances. 
Indeed, we make the point that superradiance is a useful and broad guiding
principle for  radiating systems in electrodynamics and elsewhere.

This paper is organized as follows. In Sec.~\ref{IMS}  we review
and elaborate on the Ginzburg--Frank argument for spontaneous emission in
certain modes by an object moving inertially and superluminally in a medium. 
We then give a thermodynamic argument for superradiant amplification of
those classical waves which satisfy the Ginzburg--Frank condition.  In
Sec.~\ref{examples} we discuss a number of phenomena involving amplification
of waves which can be reduced to the inertial superradiance motif; one is
a new effect.  Sec.~\ref{rotational} discusses, in the footsteps of Zel'dovich
\cite{Zeld1,Zeld2}, the thermodynamic basis of spontaneous emission and
superradiant amplification by {\it rotating\/} objects.  In
Sec.~\ref{cylinder} we give an electrodynamic proof of rotational
superradiance, and calculate in detail the superradiant gain of a conducting
(and/or otherwise dissipative) rotating cylinder for one type of
polarization.  This new treatment improves and corrects Zel'dovich's
semiquantitative estimates \cite{Zeld2}, extends the results to situations
where the material's permeabilities are not unity, and can easily be extended
to relativistic rotation (as may be relevant for pulsars).  It is followed by
a description of a device which can make rotational superradiance measurable
in the laboratory.  The Appendix extends some of the above results to the
second polarization.   

Unless otherwise noted we use units in which the speed of
light {\it in vacuum\/} is unity: $c=1$. 

\section{Inertial Motion Superradiance: Principles}  \label{IMS}
\subsection{Spontaneous Superradiance} \label{SS}

Let $E$ and $E'= E -\hbar\omega$ denote the object's total energy in the
{\it laboratory\/} frame before and after the emission of a photon with energy
$\hbar\omega$ and momentum $\hbar {\bf k}$ (both measured in the laboratory
frame), while ${\bf P}$ and  ${\bf P'}={\bf P}-\hbar{\bf k}$ denote the
corresponding momenta; ${\bf v}=\partial E/\partial {\bf P}$ is the initial 
velocity of the object. The object's rest mass $M$ is nothing but the
energy measured in the rest frame, $M=\gamma(E-{\bf v}\cdot {\bf P})$ with
$\gamma\equiv (1-{\bf v}^2)^{-1/2}$, while after the emission, with obvious
notation, $M'=\gamma'(E'-{\bf v'}\cdot {\bf P'})$.  Then a straightforward
calculation to ${\cal O}(\omega), {\cal O}({\bf k})$ and ${\cal O}({\bf
v'}-{\bf v})$ gives
\begin{equation}
M'-M=-\gamma\hbar(\omega -{\bf v}\cdot {\bf k}) +\hbar \omega \cdot {\cal
O}({\bf v'}-{\bf v})
\label{basic}
\end{equation}
As written, this formula is relevant for emission; for absorption the sign
in front of $(\omega -{\bf v}\cdot {\bf k})$ should be
reversed.   The factor ${\cal O}({\bf v'}-{\bf v})$ represents recoil effects;
it is of order $\hbar\omega/M$ and becomes negligible for a sufficiently heavy
object. In this recoilless limit
\begin{mathletters}
\label{emission}
\begin{equation}
M'-M=-\gamma\hbar(\omega -{\bf v}\cdot {\bf k}); \qquad {\rm (emission)}
\end{equation}
\begin{equation}
M'-M=\gamma\hbar(\omega -{\bf v}\cdot {\bf k}); \qquad {\rm (absorption)}
\end{equation}
\end{mathletters}

We note that in vacuum $\,\omega=|{\bf k}|>{\bf v}\cdot {\bf k}$ so that
emission is possible only with de--excitation ($M'-M<0$), while absorption is
coupled with excitation ($M'-M>0$), as plain intuition would have.

Now suppose the object moves uniformly through an isotropic medium 
transparent to electromagnetic waves possessing an index of refraction
$n(\omega)>1$.  The $\hbar\omega$ and $\hbar{\bf k}$ are still the
energy and momentum of the photon, but now
$\omega=|{\bf k}|/n(\omega)$.  Whenever
$|{\bf v}|<1/n(\omega)$ (subluminal motion for the relevant
frequency) we  recover  the connections ``de--excitation $\leftrightarrow$
emission'' and ``excitation $\leftrightarrow$ absorption''.  Ginzburg and
Frank\cite{GinzFr,Ginz} refer to this kind of emission or absorption as the
ordinary Doppler effect, because the relation between $\omega$ and ${\bf k}$
and the rest frame transition frequency
$\omega_0\equiv |M-M'|/\hbar$,
\begin{equation}
\omega_0 = \gamma(\omega - {\bf v}\cdot{\bf k}),
\label{Doppler}
\end{equation}
is the standard Doppler shift formula.

In the case $|{\bf v}|>1/n(\omega)$ the object moves faster than the {\it
phase\/} velocity of electromagnetic waves of frequency $\omega$.  If
$\vartheta$ denotes the angle between ${\bf k}$ and
${\bf v}$, a photon in a mode with $\cos\vartheta > [n(\omega)|{\bf v}|]^{-1}$
has negative $\omega -{\bf v}\cdot {\bf k}$, and can thus be {\it emitted\/}
only in consonance with  {\it excitation\/} of the object
($M'-M>0$).  Ginzburg and Frank refer to this eventuality as the
{\it anomalous\/} Doppler effect.  They note a variety of circumstances other
than superluminal motion in a dielectric for which the conditions for the
anomalous Doppler effect can be met: a particle moving in vacuum through  a
narrow channel drilled into a dielectric,  a particle shot into a gap between
dielectric slabs, emission from a collection of sources which are succesively
excited so that the active source moves along with superluminal {\it phase\/}
velocity, {\it etc.\/} \cite{GinzFr,Ginz}      

Thus an object in its ground state may become excited and emit a
photon, provided it moves superluminally through a medium.  The energy source
must be the bulk motion.  Emission is not just allowed by the conservation
laws; it will occur spontaneously, as follows from thermodynamic reasoning. 
The object in its ground state with no photon around constitutes a low entropy
state; the excitation of the object to one of a number of possible excited
states with emission of a photon with momentum in a variety of possible
directions evidently entails an increase in entropy.  Thus the emission is
favored by the second law.

Recall that according to Eq.~(\ref{emission}b), when $\omega -{\bf v}\cdot
{\bf k}< 0$ {\it absorption\/} of a photon is possible only if accompanied by
a {\it de--excitation\/} of the object ($M'-M<0$).  Thus a superluminally
moving object in the ground state is forbidden from absorbing in certain modes
!

A further case is absorption or emission by a superluminal
object of photons with the directions given precisely by $\cos\vartheta =
[n(\omega)|{\bf v}|]^{-1}$ .  According to Eqs.~(\ref{emission}) both are
possible and {\it do not\/} require a change in the object.  In fact, both
processes can occur consecutively, thus constituting scattering of a photon
with no change in the object.  Consequently, for superluminal motion,
scattering with both initial and final directions specified by
$\cos\vartheta = [n(\omega)|{\bf v}|]^{-1}$ can be coherent scattering.  In
particular, all these processes are possible for a structureless
particle which, of course, has only one state (structureless is a relative
concept; we mean the particle looks structureless at the relevant energy
scale).

\subsection{Superradiant Amplification}   \label{SA}  

The above section deals with {\it spontaneous superradiance\/} by an
elementary system.  Ginzburg \cite{Ginz}  has in mind a two level atom (a
dipole).  If the object has complicated structure, so that it may dissipate
energy internally, it is also capable of amplifying an ambient electromagnetic
wave which satisfies the superradiance condition.  We now show this by a
classical argument.

Suppose that the incident radiation is exclusively in modes with frequency
near $\omega$ and propagating within $\Delta {\bf n}$ of the direction ${\bf
n}$.  Let $I(\omega, {\bf n})$ denote the corresponding intensity (power per
unit area, unit solid angle and unit bandwidth). Experience tells us that the
body will absorb power $a(\omega, {\bf n})\,\Sigma({\bf  n})\,I(\omega,
{\bf n})\,\Delta\omega\Delta {\bf n}$, where $\,\Sigma({\bf  n})\,$ is the
geometric crossection normal to ${\bf n}$, and $a(\omega, {\bf n})<1$
is a characteristic absorptivity of the body. Simultaneously the object
will scatter power $[1-a(\omega, {\bf n})]\,\Sigma({\bf  n})\,I(\omega, {\bf
n})\,\Delta\omega\Delta {\bf n}$.  By conservation of energy  
\begin{equation}    
{dE\over dt}= a\,\Sigma\,I\,\Delta\omega\Delta {\bf
n} - W 
\label{Erate}
\end{equation}
where $W$ is the overall power spontaneously emitted by the body (including
any thermal emission).  We ignore energy going into scattered photons because
it will not show up in Eq.~(\ref{Srate}) below.  

Now the linear momentum conveyed by the radiation is ${\bf n}\, n(\omega)$
times the energy conveyed.  The easiest way to see this is to think of the
radiation as composed of quanta, each with energy
$\hbar\omega$ and  momentum
$\hbar{\bf k}$ with $\omega\,  n(\omega)  =|{\bf k}|$. However, the result also
holds classically, and can be derived, for instance, by comparing 
the temporal-spatial and spatial-spatial components of the energy-momentum
tensor for the field.  Thus absorption and spontaneous emission cause the
linear momentum ${\bf P}$ of the body to change at a rate
\begin{equation}   
{d{\bf P}\over dt}= {\bf n}\, n(\omega)\,a\,\Sigma\,I\,\Delta\omega\Delta {\bf
n} - {\bf U}  
\label{Prate} 
\end{equation}
where ${\bf U}$ signifies the rate of spontaneous momentum emission.  We
have not included the transfer of momentum due to scattering because
this has no influence on Eq.~(\ref{Srate}) below.

As already hinted, in calculating the rate of change of rest mass of the body,
$M$, we may forget the effects of elastic scattering.  For in the frame of
the body waves are scattered with no Doppler shift (since there is no
motion), which means that they contain the same energy before and after the
scattering.  Thus the scattering cannot contribute to $dM/dt$.  Because $M$
is just the body's energy in its own frame, the rest mass changes at a rate
given by a Lorentz transformation:
\begin{equation}
dM/dt = \gamma(dE/dt - {\bf v}\cdot d{\bf P}/dt)
\label{Mrate}
\end{equation}
Of course, a change in the proper mass means that the number of microstates
accessible to the object has changed, {\it i.e.\/}, that its entropy $S$ has
changed. Defining an effective temperature for the  body,
$T=\partial M/\partial S$, we see by Eqs.~(\ref{Erate})--(\ref{Prate}) that
\begin{equation} 
{dS\over dt}= {\gamma\over T}\,[\omega^{-1}\,(\omega - {\bf v}\cdot {\bf k})\,
a\,\Sigma\,I \,\Delta\omega\Delta
{\bf n} - W +{\bf v}\cdot{\bf U}\,]
\label{Srate}
\end{equation}
where we have replaced ${\bf n}\,\omega\, n(\omega) \rightarrow {\bf k}$.

Let us now take into account the rate of change of radiation entropy,
$d{\cal S}/dt$.  We get an upper bound on it by ignoring any
entropy carried {\it into\/} the object by the waves.    Now
the entropy in a single mode of a field containing on the mean $N$ quanta is at
most \cite{LLSP1} 
\begin{equation}
{\cal S}_{\rm max}= (N+1)\ln (N+1) - N\ln N \approx \ln N
\label{Smax}
\end{equation}
where the approximation applies for $N\gg 1$.   The scattered
waves carry a mean number of quanta proportional to $I(\omega, {\bf n})$.
Hence, for large $N$ , the outgoing waves' contribution to $d{\cal S}/dt$ is
bounded from above by a quantity of ${\cal O}[\ln I(\omega, {\bf n})]$.  There
is an additional contribution to $d{\cal S}/dt$ of ${\cal O}(W)$ coming from
the spontaneous emission.  Hence
\begin{equation}
d{\cal S}/dt < {\cal O}[\ln I(\omega, {\bf n})] + {\cal O}(W)
\label{Srad}
\end{equation}  

If any dissipation takes place, the second law of thermodynamics demands
$dS/dt+d{\cal S}/dt > 0$.  As $I(\omega, {\bf n})$ is made 
larger and  larger, the total entropy rate of change becomes dominated by the
term proportional to $I(\omega, {\bf n})$ in Eq.~(\ref{Srate}) because $W$ and
${\bf U}$ are kept fixed.  Positivity of $dS/dt+d{\cal S}/dt$ then
requires
\begin{equation}
(\omega - {\bf v}\cdot {\bf k})\, a(\omega, {\bf n}) > 0
\label{acondition}
\end{equation}
Thus whenever the Ginzburg--Frank condition,
\begin{equation}
\omega - {\bf v}\cdot {\bf k} < 0
\label{superradiancecondition}
\end{equation}
for the anomalous Doppler effect is fulfilled, we necessarily have $a(\omega,
{\bf n}) < 0$. This result was obtained by assuming
$a\, \Sigma\,I\,\Delta\omega\Delta {\bf n} \gg W$.  But since - barring
nonlinear effects - $a$ must be independent of the incident intensity, the
result must be true for any intensity which can still be regarded as
classical.  Now $a < 0$ means that the scattered wave, with power proportional
to $1- a$, is stronger than the incident one (which is represented by the 
``1'' in the previous expression).  Thus the moving object amplifies
preexisting radiation in modes satisfying the Ginzburg--Frank condition.  We
say that the object superradiates.  For modes with $\omega - {\bf v}\cdot {\bf
k} > 0$,
$a > 0$ and so the object absorbs on the whole.

As a rule of thumb amplification of waves may be regarded as the classical
counterpart of stimulated emission at the quantum level.  By Einstein's
argument stimulated emission goes hand in hand with spontaneous emission in
the same mode. The spontaneous emission corresponding to superradiance
amplification is just the Ginzburg--Frank emission discussed in connection
with Eqs.~(\ref{emission}).  However, the spontaneous emission coefficient is
not easily calculated from $a$; the usual Einstein relation between $A$ and
$B$ coefficients cannot be used here because the object, by virtue of its very
motion, is not in thermodynamic equilibrium with the surrounding medium.

Obviously $a$ switches sign at the superradiance treshold $\omega={\bf
v}\cdot{\bf k}$.  This switch cannot take place by $a$ having
a pole since $a <1$. {\it If\/} $a$ is
analytic in  $\omega-\Omega m$, it must thus have the expansion  
\begin{equation}
a\,=\,\alpha({\bf v}, {\bf n})\,(\omega-{\bf v}\cdot{\bf
k}) + \cdots
\label{expansion}
\end{equation}
in the vicinity of the neutral frequency $\omega\,=\,{\bf v}\cdot{\bf k}$. 
However, we must emphasize that thermodynamics does not require the
function $a$ to be continuous at $\omega\,=\,{\bf
v}\cdot{\bf k}$.   

The superradiance discussed here and in Sec.~\ref{SS} will evidently  occur
also for fields other than the electromagnetic.  All that is required is that
the energy and momentum of a quantum be expressable in terms of frequency and
wavevector in the usual way.    Thus one can replace above ``photons'' and
``electromagnetic waves'' by phonons and sound waves, {\it etc.\/}

\section{Inertial Motion Superradiance: Examples}  \label{examples}

We now give four examples of phenomena that can be understood as
manifestations of inertial motion superradiance.  One is novel. 

\subsection{Vavilov--Cherenkov Effect}  \label{Cherenkov*}

A point charge moving at speed $v$ through a transparent isotropic dielectric
medium faster than the phase speed of electromagnetic radiation for some range
of frequencies will emit radiation at all those frequencies; for each
frequency the radiation front is a cone with opening angle
$2\Theta_C(\omega)$, where
\begin{equation}
\sin\Theta_C(\omega) = [v\,n(\omega)]^{-1}
\label{Cherenkov}
\end{equation}
This Vavilov--Cherenkov effect, discovered experimentally in 1934, and
explained theoretically by Tamm and Frank \cite{LLECM}, was the
first example of coherent radiation from an unaccelerating source.  We
now elaborate on Ginzburg's \cite{Ginz} discussion of the effect in terms of
superradiance. 

Since the charge has no internal degrees of freedom, its rest
mass is fixed.  We may thus set $M'-M=0$ in Eqs.~(\ref{emission}).
Those conditions cannot thus be satisfied for $v < 1/n(\omega)$ since their
r.h.s. would then be strictly positive: no absorption or emission is
possible from a subluminal particle.  However, for $v > 1/n(\omega)$ the
r.h.s. vanishes when the photon's direction makes an angle
$\vartheta$ to the particle's velocity, where $\cos\vartheta=
[v\,n(\omega)]^{-1}$.    But then the front of photons emitted as the charge
goes by forms a cone with opening angle $2(\pi/2-\vartheta)$ which evidently 
coincides with $2\Theta_C$.  As argued in Sec.~\ref{SS}, the growth of entropy
associated with the multiplicity of possible azimuthal directions of the
emitted photon favors emission; the emitted photons constitute the
Vavilov--Cherenkov radiation.
 
In truth the above description is somewhat simplistic.  It is well known that
the Vavilov--Cherenkov radiation actually comes from those regions in the
dielectric that feel strongly the electromagnetic field of the charge
\cite{LLECM}.  In effect the charges carries along with it a polarization cloud
of dielectric material.  As the charges advances, the atomic constituents of
this cloud are replaced continuously by fresh atoms from upstream .  Because of
this renewal, the system ``charge $+$ cloud'' is a dissipative one: part of
the energy that goes into exciting an atom in the cloud is inexorably carried
away into the wake of the charge.  

The argument of Sec.~\ref{SA} then tells us that the moving charge (and its
polarization cloud) must also amplify ambient radiation which satisfies 
condition (\ref{superradiancecondition}).  Writing
${\bf v}\cdot{\bf k}= v \omega\, n(\omega) \cos\vartheta$, it follows that
amplification occurs for $\cos\vartheta >1/v\, n(\omega)$.  Thus radiation
modes inside the Cherenkov cone (those with wavevector more aligned with the
charge's motion than the Vavilov--Cherenkov modes') must be amplified.  This
Vavilov--Cherenkov superradiant amplification has not yet been observed.

\subsection{Gravitational Generation of Electromagnetic Waves}  \label{GtoEM}

We now discuss a new phenomenon.  Suppose an electrically neutral black hole
of mass $M$ moves with constant velocity ${\bf v}$  through a uniform and
isotropic dielectric with an index of refraction whose real part
is $n(\omega)$.  In order to avoid questions regarding the destructive effect
of the hole on the dielectric,  it is convenient to imagine that the dielectric
is solid, and that the hole travels down a narrow straight channel drilled
through the dielectric.  Thus the hole does not accrete material, but its
gravitational field certainly influences the dielectric. 

Let a spectrum of electromagnetic waves pervade the dielectric.  Those wave
modes for which $\omega - {\bf v}\cdot{\bf k}= \omega[1-{\bf v}\cdot {\bf
n}\,n(\omega)]<0$ can undergo superradiant amplification from the black hole. 
In the argument of Sec.~\ref{SA} the entropy of the object is now
replaced by black entropy together with entropy of the surrounding
dielectric.  Now black hole entropy is proportional to the horizon area, and 
Hawking's area theorem \cite{Hawking_area} tells us that black hole area will
increase in any classical process, such as absorption of electromagnetic waves
by the hole.   If the dielectric can dissipate, it will
also contribute to the increase in entropy through changes it undergoes in the
vicinity of the passing hole.   Thus the argument of Sec.~\ref{SA}
tells us that the black hole plus surrounding dielectric will amplify the
radiation in the mentioned modes at the expense of the hole's kinetic energy. 
Likewise, even if there are no waves to start with, the argument of
Sec.~\ref{SS} tells us that the black hole plus dielectric will spontaneusly
emit photons into modes that obey the Ginzburg--Frank superradiance condition
(\ref{superradiancecondition}).  

In the conversion of kinetic energy to waves, gravitation must obviously
play a role.  For the black hole is assumed uncharged, so that the process is
distinct from the Vavilov--Cherenkov effect.  Since the waves cannot
classically emerge from within the hole, we must look for their source in the
polarization cloud accompanying the hole.  This cloud forms because  gravity
pulls on the positively charged nuclei in the dielectric stronger than on the
enveloping electrons.  As a result the array of nuclei sags with respect to
the electrons, and produces an electrical polarization of the dielectric
accompanied by an electric field which ultimately balances the tendency of
gravity to rip out nuclei from electrons.  It is this electric structure which
is to be viewed as the true source of the photons.

In special circumstances the present problem may be mapped onto that of the
Vavilov--Cherenkov effect by noting that the induced electric field
${\bf E}$ is related to the gravitational one, ${\bf g}$, by $e{\bf E}=-\delta
\mu\, {\bf g}$ where $\delta \mu\approx A m_p$ is the nucleus--electron mass
difference ($A$ is the mass number of the atoms, $m_p$ the
proton's mass), and $e>0$ the unit of charge.  From the gravitational
Poisson equation it follows that $\nabla\cdot{\bf E}=4\pi G M(\delta
\mu/e)\delta({\bf r}-{\bf r}_0)$ where
${\bf r}_0$ denotes the momentary black hole position.  The electric field
accompanying the black hole is thus that of a pointlike charge
$Q\equiv GAMm_p/e$.  There is a big assumption here that the dielectric has
time to relax to form the above compensating field.  Such relaxation does occur
for sufficiently small $|{\bf v}|$, but since we need $|{\bf v}|$ to be
sufficiently large for the Ginzburg--Frank condition to hold, stringent
conditions are required of the dielectric (high $n$ and fast relaxation). 
When these are satisfied the electromagnetic radiation will be of the
Vavilov--Cherenkov form for the equivalent  charge $Q$ moving with velocity
${\bf v}$.  $Q/e$ is about $10^3A$ times the gravitational radius of the
hole measured in units of the classical radius of the {\it electron\/}.  Hence
a fast $10^{15}$ g primordial black hole moving in a suitable dielectric 
would radiate just like an equally fast particle bearing  $\sim 10^3 A$
elementary charges.  This is relevant for the experimental search for
primordial black holes.

When things are looked at this way, the black hole character of the object is
not critical.  What matters is that it is endowed with a gravitational field.
This tells us that an ordinary object with the same mass would have similar
effect as a black hole, so long as both are smaller than the channel's
width.   It is also worthwhile noting that the effects here discussed will be
significant only when the wavelengths involved are large compared to the
width of the channel.  Otherwise, the object acts as if in vacuum, and we
expect no superradiance.

\subsection{Critical Speed for Superfluidity}  \label{Superfluids}

A superfluid can flow through thin channels with no friction.  However, when
the speed of flow is too large, the superfluidity is destroyed.  Landau gave
a criterion \cite{LLSP2} for the critical speed $v_c$ for removal of
superfluidity.  Although in practice superfluidity disappears already at much
lower speeds as the superfluid develops turbulence through the formation of
vortices, the Landau critical speed is the top speed at which superfluidity
can survive no matter how carefully tailored the channel is to the flow. The
Landau critical speed is
\begin{equation}
v_c \equiv {\rm min}\ \varepsilon({\bf p})/|{\bf p}|
\label{critical}
\end{equation}
where $\varepsilon({\bf p})$ is the dispersion relation of the
quasiparticles (phonons and rotons) that can occur as excitations above the
condensate constituting the superfluid.  In superfluid He$^4$
$v_c\approx 6\times  10^3$ cm s$^{-1}$. Landau's argument is that at speeds of
flow above $v_c$ it becomes energetically permissible for bulk kinetic
energy of the superfluid to transform into energy of one internal excitation -
a quasiparticle.  Once an abundance of quasiparticles has appeared, there is a
normal component to the fluid, which,  of course, is not a superfluid.

The Landau argument is usually framed in the rest frame of the fluid
with respect to which the walls of the channel are in motion \cite{LLSP2}. 
In the following argument we also employ that frame.  Now the walls play the
role of the object in our superradiance argument, and the waves of frequency
$\omega=\varepsilon/\hbar$ and wavenumber ${\bf k}={\bf p}/\hbar$
associated with the quasiparticles, are surrogates of the
electromagnetic waves in the arguments of Sec.~\ref{IMS}.  When the walls move
with speed  $>v_c
\equiv {\rm min}\ \varepsilon({\bf p})/|{\bf p}|$, the quantity
$\omega - {\bf v}\cdot{\bf k}= (\varepsilon - {\bf v}\cdot{\bf p})/\hbar\ $
becomes negative for at least one quasiparticle mode. It then becomes
entropically preferable for the wall material to become excited {\it
and\/} simultaneously create a quasiparticle in that mode, as discussed in
Sec.~\ref{SS}.  Furthermore, the quasiparticles thus created can undergo
superradiant amplification upon impinging on other parts of the walls
(Sec.~\ref{SA}).  As a consequence an avalanche of quasiparticle formation
ensues, which acts to convert the superfluid into a  normal fluid.  It is
clear that the transition away from superfluidity is a literal  example of the
superradiance phenomenon.  In this phenomenon the sound speed, of order
$v_c$, plays the role of the speed of light in our original arguments.  

\subsection{Superradiance in Mach Shocks}  \label{Mach}

It is well known that when a solid object travels through an
originally quiescent fluid with a speed $v=|{\bf v}|$ exceeding that of sound
$c_s$ in the fluid, a shock (density discontinuity) in the form of a circular
cone is formed in its wake \cite{LLFM}.  The interior of this Mach cone is
filled by perturbations originating in the object, while the fluid exterior to
the cone is still unperturbed.  The opening angle of the cone, $2\Theta_M$, is
easily determined by considering the locus of sound signals emitted by the
object and traveling in all directions at speed $c_s$ with respect to the
fluid which convects them downstream \cite{LLFM}:
\begin{equation}
\sin \Theta_M = c_s/v; \qquad 0 <\Theta_M < \pi/2
\label{Mach_angle}
\end{equation}
The cone's opening angle is the same in both the object's and the fluid's rest
frames.

Let us look at Mach shocks from the vantage point of superradiance.  In the
rest frame of the fluid, the object - considered structureless - can emit
{\it phonons\/} spontaneously when these satisfy the Ginzburg--Frank condition
in the form
\begin{equation}
\omega - {\bf v}\cdot{\bf k} = \omega - v k \cos \vartheta = 0.
\label{shockcriterion}
\end{equation}
Now for phonons $\omega = c_s\, k$; hence they are
spontaneously emitted at an angle $\vartheta$ to the object's velocity ${\bf
v}$ such that $\cos \vartheta = c_s/v$.  These phonons thus have components
of velocity $c_s\sqrt{1-c_s^2/v^2}$ and $c_s^2/v$ normal and parallel to ${\bf
v}$, respectively.  A Galilean transformation (velocity ${\bf v}$) to the rest
frame of the object gives for the angle $\vartheta'$ of superradiance emission
in the new frame
\begin{equation}
\sin\vartheta' = {c_s\sqrt{1-c_s^2/v^2}\over\left[(c_s\sqrt{1-c_s^2/v^2})^2 +
(c_s^2/v- v)^2\right]^{1/2}} = {c_s\over v}; \qquad \pi/2 <\vartheta' < \pi
\label{newtheta}
\end{equation}
The range of $\vartheta'$ is so chosen because in the new frame the component
of phonon velocity collinear with the object's velocity, $c_s^2/v- v$, is
negative indicating that the emission occurs into the back hemisphere, that
containing the fluid's velocity.  Because $\sin \vartheta' = \sin\Theta_M$  
we conclude that the superradiant phonons  are emitted from the object {\it
along\/} the shock discontinuity.

Now as the shock follows the object with velocity ${\bf v}$,  it advances
normal to itself with speed $v\cdot \sin\Theta_M=c_s$.  According to shock
theory \cite{LLFM}, a shock with speed $c_s$ is a weak discontinuity, {\it
i.e.\/} the fluid's density is nearly the same on both its sides.  It thus
seems possible that the shock itself is entirely made up of superradiant
phonons.

Further, consider any sound waves, {\it e.g.\/} thermal phonons,  present in
the fluid before the arrival of the object. The object is - by assumption -
structureless; however, it is accompanied in its motion by a boundary layer of
fluid that partially ``sticks'' to it \cite{LLFM}.  Because  the layer  is
constantly being renewed as the ``old'' fluid in it is swept downstream, it is
dissipative.  Therefore, those waves which satisfy the Ginzburg--Frank
condition (\ref{superradiancecondition})  will be amplified as they are
overtaken by the object.  These waves propagate at angles $\vartheta$ to the
object's direction which obey
\begin{equation}
\cos\vartheta > \omega(|{\bf k}||{\bf v}|)^{-1} = c_s/v
\end{equation}
{\it i.\,e.\/} they are emitted inside the Mach cone.  In addition, if we
regard the object with its boundary layer as one with many possible
energy states, then phonons can be emitted also by Ginzburg and Frank's
anomalous Doppler emission (see Sec.~\ref{SA}).  These also travel inside the
Mach cone.  Thus the entire acoustic ``noise'' originating from supersonic
motion in a fluid has a superradiance interpretation.

\section{ROTATIONAL SUPERRADIANCE: PRINCIPLES}\label{rotational}

We focus on an axisymmetric {\it macroscopic\/} body rotating rigidly with
constant angular velocity $\Omega$ about its symmetry axis which is supposed
fixed. The assumption of axisymmetry is critical; otherwise precession of the
axis would arise.   We further assume the body contains many internal degrees
of freedom, so that it can internally dissipate absorbed energy.  We assume it
has reached internal equilibrium and has well defined entropy $S$, rest mass
$M$ and temperature $T$. 

The body is exposed to external radiation {\it in vacuum\/}.  We classify the
radiation modes by frequency $\omega$ and azimuthal number $m$.  This last
refers to the axis of rotation.  Suppose that in the modes with azimuthal
number $m$ and frequencies in the range in $\{\omega, \omega +
\Delta\omega\}$,  power
$I_m(\omega)\,\Delta\omega$ is incident on the body.   Then, as is easy to
verify from the energy--momentum tensor, or from the quantum picture of
radiation, the radiative angular momentum is incident at rate
$(m/\omega)I_m(\omega)\,\Delta\omega$.  If  $I_m(\omega)$ is
large enough, we can think of the radiation as classical.  Experience tells us
that the body will absorb a fraction $a_m(\omega)$ of the incident
power and angular momentum flow in the modes in question, where
$a_m(\omega)<1$ is a characteristic coefficient of the body. A fraction
$[1-a_m(\omega)\,]$  will be scattered into modes with the same $\omega$ and
$m$.   We may thus replace Eqs.~(\ref{Erate})-(\ref{Prate}) by
\begin{equation}    
{dE\over dt}= a_m\,I_m\,\Delta\omega - W \label{newErate}
\end{equation}
and
\begin{equation}   
{dJ\over dt}= (m/\omega)\,a_m\,I_m\,\Delta\omega - U_J 
\label{Jrate} 
\end{equation}
where $J$ is the body's angular momentum and $U_J$ is the overall rate of
spontaneous angular momentum emission in waves.

Now the energy $\Delta E_0$ of a small system measured in a frame rotating
with angular frequency ${\bf \Omega}$ is related to its energy $\Delta E$ and
angular momentum $\Delta{\bf J}$ in the inertial frame by \cite{LLMech}
\begin{equation}
\Delta E_0 = \Delta E - {\bf \Omega}\cdot\Delta{\bf J}
\label{restenergy}
\end{equation}
Thus, when as a result of interaction with the radiation, the energy of
our rotating body changes by $dE/dt\times \Delta t$ and its angular momentum
in the direction of the rotation axis by $dJ/dt\times \Delta t$,  its
rest mass--energy changes by $(dE/dt - \Omega dJ/dt)\times
\Delta t$.  From this we infer, in parallel to the derivation of
Eq.~(\ref{Srate}), that the body's entropy changes at a rate
\begin{equation} 
{dS\over dt}= {1\over T}\left[\,{\omega - m \Omega\over
\omega}\,a_m\,I_m\,\Delta\omega - W + \Omega\,U_J\, \right]
\label{newSrate}
\end{equation}

As in the discussion involving Eqs.~(\ref{Smax})-(\ref{Srad}) we would now
argue that when $I_m(\omega)$ is large, the term proportional to $(\omega - m
\Omega)\,a_m(\omega)$ in Eq.~(\ref{newSrate}) dominates the overall
entropy balance.   The second law thus demands that
\begin{equation}
(\omega - m \Omega)\,a_m(\omega) > 0
\label{newcondition}
\end{equation}
It follows that whenever the Zel'dovich - Misner condition
(\ref{supercriterion}) is met, $a_m(\omega) <0$ necessarily.  As in
Sec.~\ref{SS}, we can argue that the sign of $a_m(\omega)$ should not depend
on the strength of the incident radiation if nonlinear radiative effects do
not intervene.  Hence, independent of the strength of $I_m(\omega)$, condition
(\ref{supercriterion}) is the generic condition for rotational superradiance.

Evidently $a_m(\omega)$ switches sign at $\omega=\Omega m$.  This
switch cannot take place by $a_m(\omega)$ having a pole there since 
$a_m(\omega)<1$. {\it If\/} $a_m(\omega)$ is analytic in  $\omega-\Omega
m$, it must thus have the expansion  
\begin{equation}
a_m(\omega)=\alpha_m(\Omega)\,(\omega-\Omega m)+\cdots
\label{expansion2}
\end{equation}
in the vicinity of $\omega=\Omega m$.  However, we must again stress
that thermodynamics does not demand continuity of $a_m(\omega)$ at
$\omega-\Omega m=0$.  Specific examples like that of the rotating
cylinder [(Eq.~(\ref{am}) below] do show continuity. 

\section{Superradiance of a Rotating Cylinder}  \label{cylinder}

Devices for making rotational superradiance observable (see
Sec.~\ref{devices} below) are modeled on Zel'dovich's rotating
cylinder \cite{Zeld2}.  In this section we idealize the cylinder as infinitely
long.  Let its radius be $R$ and let it be rotating rigidly in vacuum  with
constant angular frequency $\Omega$.  We suppose it to be made of
material with {\it spatially uniform\/} permittivity $\epsilon(\omega)$ and
permeability $\mu(\omega)$; these are not necessarily real because of the
possibility of dissipative processes in the material.  Alternatively, the
material may be electrically conducting in which case we denote its
conductivity by $\sigma$.  Although it is possible to represent conductivity as
an imaginary part of $\epsilon(\omega)$, we shall not do so here.  If $\sigma$
is small, {\it e.g.\/} a semiconductor, one can allow nontrivial
$\epsilon(\omega)$  and $\mu(\omega)$ alongside $\sigma$. 

\subsection{Constitutive Relations and Maxwell Equations}\label{equations}

In the relativistic treatment we have in mind the electromagnetic field
is described by the antisymmetric tensor $F^{\alpha\beta}$ composed in the
usual way of the electric field ${\bf E}$ and magnetic induction ${\bf B}$. 
The electric displacement ${\bf D}$ and magnetic field ${\bf H}$ form an
analogous tensor  $H^{\alpha\beta}$.  The usual constitutive relations
${\bf D}=\epsilon {\bf E}$, ${\bf B}=\mu {\bf H}$ and ${\bf j} = \sigma
{\bf E}$ can be expressed in covariant form as
\begin{mathletters}
\label{constitutive}
\begin{equation}
H^{\alpha\beta} u_\beta = \epsilon F^{\alpha\beta} u_\beta
\end{equation}
\begin{equation}
^*F^{\alpha\beta} u_\beta = \mu ^*H^{\alpha\beta} u_\beta \label{equationb}
\end{equation}
\begin{equation}
j^\alpha = \sigma F^{\alpha\beta} u_\beta + \varrho u^\alpha
\end{equation}
\end{mathletters}
We have written the electric current as a sum of a conductive part (recall
that electric and  magnetic fields are observer dependent concepts and are here
computed in the  frame of the material whose 4-velocity is $u^\alpha$) and a
convective part with $\varrho$ being the {\it proper\/} charge density. 
This last is included to give us the flexibility to treat, say, a dielectric
bearing a net charge density (in which case we would set $\sigma=0$).
We use the notation $^*F^{\alpha\beta}\equiv {1\over
2} \varepsilon^{\alpha\beta\gamma\delta}F_{\gamma\delta}$ with 
$\varepsilon^{\alpha\beta\gamma\delta}$ the Levi--Civita
tensor.  It should be observed that $\epsilon$ and $\mu$ are frequency
dependent in general, so that equations involving them refer to time Fourier
components of fields.  And the arguments of $\epsilon$ or
$\mu$ should be frequencies {\it in the frame of the rotating cylinder.\/}  

In cylindrical coordinates $\{x^0, x^1, x^2,
x^3\}=\{t, r, \phi, z\}$ with flat metric
\begin{equation}
ds^2 = -dt^2 +dr^2 + r^2 d\phi^2 + dz^2
\label{metric} 
\end{equation}
we obviously have inside the cylinder
\begin{equation}
u_\beta = (-1, 0, \Omega r^2, 0)\gamma; \quad \gamma\equiv (1-\Omega^2
r^2)^{-1/2} \label{gamma}
\end{equation}
It is easy to generalize this to curved spacetime, but we shall
not do so here.

By succesively taking $\alpha=0, 1, 3$ in Eqs.~(\ref{constitutive}a,b) and
converting components of duals to components of the original fields we get
\begin{mathletters}
\label{algebraic}
\begin{equation}
\epsilon^{-1} H^{02}= F^{02} \equiv r^{-1} E_\phi
\end{equation}
\begin{equation}
 \mu H^{31} =  F^{31} \equiv B_\phi
\end{equation}
\begin{equation}
\epsilon^{-1}(H^{01}+\Omega r^2 H^{12})= F^{01}+\Omega r^2 F^{12}
\equiv \gamma^{-1} E_r
\end{equation}
\begin{equation}
\mu (H^{23}-\Omega H^{03}) = F^{23}-\Omega  F^{03}
\equiv (r\gamma)^{-1} B_r
\end{equation}
\begin{equation}
\epsilon^{-1}(H^{03}-\Omega r^2 H^{23})= F^{03}-\Omega r^2 F^{23}
\equiv  \gamma^{-1} E_z
\end{equation}
\begin{equation}
\mu (H^{12}+\Omega H^{01}) = F^{12}+\Omega  F^{01} 
\equiv (r\gamma)^{-1} B_z
\label{equationf}
\end{equation}
\end{mathletters}
Here $E_r, E_\phi, E_z, B_r$, $B_\phi$ and $B_z$ denote the physical
components in the indicated directions of the electric field and magnetic
induction {\it in the rotating frame\/}.  Outside the cylinder one should set
$\Omega=0$ and $\epsilon=\mu=1$ in these equations. 

Let us now pass to the Maxwell equations:
\begin{mathletters}
\label{Maxwell}
\begin{equation}
F_{[\alpha\beta,\gamma]} = 0
\end{equation}
\begin{equation}
H^{\alpha\beta},_\beta = 4\pi j^\alpha 
\end{equation}
\end{mathletters}
In view of the symmetries of the problem we shall look for solutions where the
fields vary as $f(r)e^{\imath(m\phi+kz-\omega t)}$ with $m$ an integer, and
$\omega$ and $k$ real constants.  Here $\omega$ is the frequency in the
laboratory frame; in the cylinder's (rotating) frame, the azimuthal coordinate is
$\phi'=\phi-\Omega t$, and hence the frequency is $\omega'=\omega-m\Omega$. 
Our choice of modes means that in writing the equations one can simply
replace $\partial/\partial\phi\rightarrow \imath m$, {\it etc.}  From
Eq.~(\ref{Maxwell}a) we get, after raising indeces,
\begin{mathletters}
\label{newhomogeneous}
\begin{equation}
\partial(F^{02} r^2)/\partial r -\imath\omega r^2 F^{12} -\imath m F^{01} =0 
\end{equation}
\begin{equation}
\partial(F^{23} r^2)/\partial r +\imath k r^2 F^{12} +\imath m F^{31} =0
\end{equation}
\begin{equation}
\partial  F^{03}/\partial r +\imath\omega F^{31} - \imath k F^{01} = 0
\end{equation}
\begin{equation}
\imath k F^{02} +\imath\omega F^{23} -\imath m r^{-2} F^{03} = 0
\end{equation}
\end{mathletters}

Finally we take in Eq.~(\ref{Maxwell}b) succesively $\alpha=0,1,2$ and 3:
\begin{mathletters}
\label{source}
\begin{equation}
\partial(H^{01} r)/\partial r +\imath m r H^{02} +\imath k r H^{03} =
4\pi\sigma\gamma\Omega r^2 E_\phi +  r\gamma\varrho
\end{equation}
\begin{equation}
\imath\omega H^{01} +\imath m H^{12} - \imath k H^{13} = 4\pi\sigma E_r 
\end{equation}
\begin{equation}
\partial(H^{12} r)/\partial r -\imath \omega r H^{02} -\imath k r H^{23} =
-4\pi\sigma\gamma E_\phi -r\gamma\varrho\Omega
\end{equation}
\begin{equation}
\partial(H^{31} r)/\partial r -\imath m r H^{23} +\imath\omega r H^{03} =
4\pi\sigma E_z 
\end{equation}
\end{mathletters}
Outside the cylinder one should put $\sigma=0$ and $\varrho=0$ in
Eqs.~(\ref{source}).

\subsection{Axial Electric and Magnetic Modes ($k=0$)}\label{modes}

As in any electromagnetic problem of this type, there are here two distinct
modes for each set $\{\omega, m, k\}$.  Here we characterize them for the case
$k=0$.

First assume, in harmony with Eq.~(\ref{algebraic}a) that everywhere
inside and outside the cylinder $F^{02}=H^{02}=E_\phi=0$.  It will transpire
that this is a consistent choice, and therefore characterizes the first mode.
 Eq.~(\ref{newhomogeneous}a) then gives $\omega r^2 F^{12}+ m F^{01}=0$
everywhere, while outside the cylinder
($H^{\alpha\beta}=F^{\alpha\beta}$; $\sigma=\varrho=0$) Eq.~(\ref{source}b)
gives $\omega F^{01} + m F^{12} =0$. These simultaneous equations require
$F^{12}=F^{01}=H^{12}=H^{01}=0$ outside the cylinder. To connect these with
the interior fields we go to Eqs.~(\ref{source}a,c).  There may be a charge
layer at $r=R$ of  surface density $\Delta q=\int_{R-}^{R+} \varrho dr$. 
Integrating the two equations across the layer gives for the jumps in the
fields $\Delta H^{01} = q\gamma(R)$ and  $\Delta H^{12} = -q\Omega
\gamma(R)$ so that $\Omega H^{01}+H^{12}$ must be  continuous
across the surface.  If we now add $\Omega$ times Eq.~(\ref{source}a) to
Eq.~(\ref{source}c) we find that $r(\Omega H^{01}+H^{12})$ is independent of
$r$ everywhere, including at $r=R$.  Since it vanishes for $r>0$, it must
vanish everywhere.  Then by Eq.~(\ref{algebraic}f) $\Omega F^{01}+F^{12}=0$
everywhere.  But as we mentioned, $\omega r^2 F^{12}+ m F^{01}=0$ everywhere;
these two simultaneous equations force $F^{01}$ and $F^{12}$ to vanish
everywhere.  It is now evident by solving Eqs.~(\ref{algebraic}c,f)
simultaneously that $H^{01}$ and $H^{12}$ must also vanish everywhere. 

As we shall show in Sec.~\ref{flux}, one can construct $F^{31}$ and $F^{23}$
out of $F^{03}$ which obeys an autonomous equation.  Thus the ansatz
$F^{01}=F^{02}=F^{12}=H^{01}=H^{02}=H^{12}=0$ defines a 
mode of the system.  We call it the axial electric (AE) mode because its
electric field (only component $F^{03}$) points along the cylinder's axis.  It
corresponds to Zel'dovich's \cite{Zeld2} first mode.

Now we look for a mode which has [see Eq.~(\ref{algebraic}b)]
$F^{31}=H^{31}=B_\phi=0$.  Again, it will transpire that this is a consistent
choice.  From Eqs.(\ref{newhomogeneous}b,c) it follows that $r^2 F^{23} = C_1$
and $F^{03}= C_2$ with $C_1$ and $C_2$ independent of $r$. 
Eq.~(\ref{source}d) implies that outside the cylinder $mF^{23}-\omega
F^{03}=0$.  This last is inconsistent with the previous expressions unless we
put $C_1=C_2=F^{03}=F^{23}=H^{03}=H^{23}=0$ in the exterior.  Now since 
$F^{23}$ is the magnetic field component  {\it normal\/} to the cylinder's
surface, it must be continuous there.  Thus $C_1$ along with $F^{23}$
must also vanish inside the cylinder.  The {\it tangential\/} electric field
$F^{03}$ must likewise be continuous at the surface; thus $C_2$ and $F^{03}$
have to vanish inside as well.  By solving Eqs.~(\ref{algebraic}d,e)
simultaneously we find that also
$H^{03}=H^{23}=0$ inside.

As we show in the Appendix, one can construct $F^{02}, F^{12}$
and $F^{01}$ out of a single function obeying an autonomous equation.  Thus
the ansatz $F^{03}=F^{31}=F^{23}=H^{03}=H^{31}=H^{23}=0$ defines a second
mode.  We call it the axial magnetic (AM) mode because its magnetic field
(only component $H^{12}$) points along the cylinder's axis.  It corresponds to
Zel'dovich's \cite{Zeld2} second mode.

\subsection{Electrodynamic Proof of Superradiance for AE Modes
($k=0$)}\label{flux}

Here we give a new basically electrodynamic proof that for $\omega-m\Omega<0$
the cylinder superradiates.  We shall first obtain the radial equation
governing the shape of the AE mode with  $k=0$.  First we note that according
to Eqs.~(\ref{newhomogeneous}c,d), 
\begin{mathletters}
\label{true}
\begin{equation}
F^{31} = \imath\omega^{-1}  \partial F^{03}/\partial r
\end{equation}
\begin{equation}
F^{23} = m\omega^{-1} r^{-2} F^{03}
\end{equation}
\end{mathletters}
Next we solve for $H^{03}, H^{23}$ and $H^{31}$ from
Eqs.~(\ref{algebraic}b,d,e) and substitute these and Eq.~(\ref{true}a) in
Eq.~(\ref{source}d) to get
\begin{equation}
r^{-1}\partial(r\partial F^{03}/\partial r)/\partial r
+m\omega\gamma(B_r+\epsilon\mu\Omega E_z)-\omega^2\gamma(\epsilon\mu
E_z+\Omega r^2B_r) = 4\pi\imath\sigma\mu E_z
\label{radial1}
\end{equation}
But by combining the definitions of $E_z$ and $B_r$ in Eqs.~(\ref{algebraic})
with Eq.~(\ref{true}b) we have that
\begin{mathletters}
\label{E/B}
\begin{equation}
E_z = \gamma\omega^{-1}(\omega-m\Omega) F^{03}
\end{equation}
\begin{equation}
B_r =\gamma(\omega r^2)^{-1} (m-\omega\Omega r^2) F^{03}
\end{equation}
\end{mathletters}
If we now substitute these in Eq.~(\ref{radial1})  and cancel out the
common phase $e^{\imath(m\phi-\omega t)}$ we get
\begin{equation}
r^2 f'' + rf' -\gamma^2[(m-\omega\Omega r^2)^2-\epsilon\mu
(\omega-m\Omega)^2 r^2-4\pi\imath\gamma^{-1}\mu\sigma(\omega-m\Omega)r^2]f=0
\label{radial2}
\end{equation}
where $f(r)\equiv F^{03}e^{-\imath(m\phi-\omega t)}$ and $'$ denotes an
ordinary radial derivative.  All this is for $r<R$.  In the cylinder's
exterior we just set $\epsilon\mu\rightarrow 1$ and $\sigma\rightarrow 0$.
 This is the promised exact radial equation for the AE mode; the fields
$F^{31}$ and $F^{23}$ can be recovered from Eqs.~(\ref{true}).

Now we are ready to discuss the energy flux.  Both inside and outside the
cylinder the {\it radial} energy flux is \cite{LLECM} 
\begin{equation}
S_r = ({\bf E}\times {\bf H})_r/4\pi= (F^{02} H^{12} -F^{03}
H^{31})/4\pi
\label{radialflux}
\end{equation}
But $F^{02}$ and $H^{12}$ both vanish, so this reduces to 
$-F^{03} H^{31}/4\pi$.
This is the instantaneous flux; of more interest is the time--averaged flux
which can be obtained by substituting \cite{LLECM}
\begin{mathletters}
\label{realfields}
\begin{equation}
F^{03} \rightarrow {1\over 2}\left[f e^{\imath(m\phi-\omega t)} +
f^*e^{-\imath(m\phi-\omega t)}\right]
\end{equation}
\begin{equation}
H^{31} = F^{31}/\mu \rightarrow{1\over 2\omega}\left[\imath(f'/\mu)
e^{\imath(m\phi-\omega t)} -\imath (f^{* '}/\mu^*) e^{-\imath(m\phi-\omega
t)}\right]
\end{equation}
\end{mathletters}
and then averaging.  Here we have used Eq.~(\ref{true}a) to simplify.  Note
that the complex conjugate of the primary field $f$ contributes with weight
$1/\mu^*$.  We thus have for the time--averaged radial flux
\begin{equation}
\overline S_r ={1\over 16\pi\imath\omega}\left(f^* f'/\mu - f
f^{*'}/\mu^*\right)
\label{energyflux}
\end{equation}
In the process two terms involving exponents $e^{\pm 2\imath(m\phi-\omega t)}$
have averaged out.

We can get a useful equation for the Wronskian--like expression in the last
equation by first dividing Eq.~(\ref{radial2}) by $r\mu$, multiplying it by
$f^*$, and then substracting from the result its complex conjugate:
\begin{equation}
{d\over dr}[r(f^* f'/\mu - f f^{*'}/\mu^*)] = -2\imath r[{\cal A}|f'|^2+
({\cal B+C}) |f|^2 ]
\label{Wronskian}
\end{equation}
with ($\Im$ means imaginary part)
\begin{mathletters}
\label{dissipfactor}
\begin{equation}
{\cal A}\equiv \Im\mu/|\mu|^{2}
\end{equation}
\begin{equation}
{\cal B} \equiv [\Im\epsilon(\omega-m\Omega)^2 +
{\cal A}(m-\omega\Omega r^2)^2r^{-2}]\gamma^2
\end{equation}
\begin{equation}
{\cal C}\equiv 4\pi\sigma(\omega-m\Omega)\gamma
\end{equation}
\end{mathletters}
In the vacuum outside the cylinder $\Im\epsilon=\Im\mu=\sigma=0$ so that
according to Eqs.~(\ref{energyflux}-\ref{Wronskian}) $S_r\propto 1/r$.  This
just means that energy is conserved outside the cylinder, the overall outflow
(inflow) at large distances equaling that at $r=R$.  Thus to find out which
way energy flows at large distances, it is sufficient to determine the sign of
$\overline S_r$ at $r=R$.

Now because $f$ represents a physical electric field, it must be bounded at
$r=0$.  And then $f'$ cannot diverge as fast as $1/r$.  It follows that,
barring the exceptional circumstance that $\mu=0$,  $r(f^* f'/\mu - f
f^{*'}/\mu^*)\rightarrow 0$ as $r\rightarrow 0$.  Hence by integrating
Eq.~(\ref{Wronskian}) from $r=0$ to $r=R$ we find
\begin{equation}
\overline S_r(r=R) ={-1\over 32\pi\omega R}\int_0^R r[{\cal A}|f'|^2+({\cal
B+C}) |f|^2 ] dr
\label{flux2}
\end{equation}
To determine the sign of this expression we note that it follows from the
second law of thermodynamics \cite{LLECM} that $\sigma\geq 0$, and
that $\Im\epsilon$ and $\Im\mu$ are both odd in the frequency and both 
positive for positive frequency.  Of course, frequency here means frequency in
the frame of the material, namely $\omega-m\Omega$.  Hence ${\cal A}, {\cal
B}$ and ${\cal C}$ all bear the same sign as $\omega-m \Omega$.  Thus
regardless of the source of dissipation, there is an energy outflow to infinity
(superradiance) if only if the Misner--Zel'dovich condition
$\omega-m\Omega<0$ is satisfied, as we might have guessed from the method of
Sec.~\ref{rotational}.

\subsection{Gain in Superradiance for Nonrelativistic
Rotation: AE Modes}\label{amplification}  

In his pioneering study of superradiance of a rotating cylinder,  Zel'dovich
\cite{Zeld2} concluded that for AE modes with $\omega-m\Omega<0$, 
$m>0$ and $k=0$,  the gain coefficient
[defined precisely after Eq.~(\ref{y})]  is very small for
nonrelativistic rotation. The gist of his argument is as follows. 
Outside the cylinder the radial equation~(\ref{radial2}) reduces exactly to
\begin{equation}
r^2 f'' + rf' -[m^2-\omega^2 r^2]f=0; \quad r> R
\label{radialoutside}
\end{equation}
whose solutions are the Hankel functions $H_m^{(1)}(\omega r)$ and
$H_m^{(2)}(\omega r)$, the first (second)  representing outgoing (ingoing) 
waves at infinity.  Inside the cylinder Zel'dovich takes $\epsilon= 
\mu=1$, and neglects the effect of $\sigma$ to argue that one may, to
sufficient accuracy, approximate $f$ by $J_m(\omega r)$ which is that
combination of $H_m^{(1)}(\omega r)$ and $H_m^{(2)}(\omega r)$ 
regular at $r=0$.  We may justify this form by realizing that
\begin{equation}
[(m-\omega\Omega r^2)^2- (\omega-m\Omega)^2 r^2]\gamma^2 =
(m^2-\omega^2 r^2)
\label{fuchs}
\end{equation}
so that in the stated limit Eq.~(\ref{radial2}) reduces to
Eq.~(\ref{radialoutside}) also inside the medium.  This is true even for
relativistic rotation, a point not remarked on by Zel'dovich.

Working nonrelativistically Zel'dovich then calculates via Ohm's law the
current ${\bf j}$ induced in the cylinder by the electric and magnetic fields 
${\bf E}$ and  ${\bf B}$ deriving from this $f$.  Because the medium rotates,
he finds that $j_z\propto (\omega-m\Omega)$.  Thus the Joule work $j_z E_z$ is
negative: the cylinder does work on the field and superradiance ensues. 
Zel'dovich obtains a gain coefficient $\propto \sigma\cdot (m\Omega-m)(\omega
R)^{2m}$.  The factors $\omega R$ come from the small argument approximation
$J_m(x) \sim x^m$ for $x\ll m$; recall that because of the Zel'dovich--Misner
condition and the assumed nonrelativistic rotation, $\omega r <m\Omega R\ll
m$.  As Zel'dovich remarks, the physical reason for the smallness is that $R$
lies deep within the near zone,  which circumstance suppresses the
matter--wave coupling.

Is Zel'dovich's pessimistic conclusion valid also when $\epsilon, \mu\neq
1$ ?  One may be skeptic because when $\epsilon\mu$ differs
significantly from unity, Eq.~(\ref{radial2}) does not reduce to
Eq.~(\ref{radialoutside}), but rather to the Bessel equation
(\ref{inside2}a) below whose solution regular at $r=0$ is different
from $J_m(\omega r)$.   One also wonders what happens when the conductivity is
large, so that the backreaction of the cylinder on the wave cannot be
neglected, and when $\gamma$ is significantly greater than unity ?   To answer
these question we shall work with the full Eq.~(\ref{radial2}), and match its
interior and exterior solutions.  We can then be more specific about the
prefactor in Zel'dovich's expression and the corrections it is subject to
for large $\sigma$.

Let us assume that the ingoing wave generated by some external agency,
$H_m^{(2)}(\omega r)$, has unit coefficient.  Then the total radial wave
amplitude outside the cylinder will be $f_{\rm out}= H_m^{(2)}(\omega r)+\rho
H_m^{(1)}(\omega r)$ where $\rho$ is the (possibly complex) amplitude for
reflection off the cylinder.  For superradiance we expect $|\rho|^2>1$.  

Inside the cylinder the exact $f(r)$ is determined by Eq.~(\ref{radial2})
which in light of Eq.~(\ref{fuchs}) can be rewritten in the more convenient
form
\begin{mathletters}
\label{inside2}
\begin{equation}
r^2 f'' + rf' -[m^2-\kappa^2 r^2]f=0; \quad r< R
\end{equation}
\begin{equation}
\kappa^2 \equiv \omega^2 +(1-\epsilon\mu)(\omega-m\Omega)^2 \gamma^2 + \imath\,
4\pi\gamma\mu\sigma(\omega-m\Omega)
\end{equation}
\end{mathletters}
This is again a Bessel equation whose solution regular at
$r=0$ is $J_m(\kappa r)$.  The radial wave amplitude inside will thus be
$f_{\rm in}=\tau J_m(\kappa r)$ where $\tau$ is the (possibly complex)
amplitude for transmission into the cylinder.

Now we match interior with exterior solutions by the usual
continuity conditions on electric and magnetic fields.  By integrating
Eq.~(\ref{true}a) from $r=R-\varepsilon$ to $r=R+\varepsilon$ and relying on
the boundedness of $F^{31}$ we conclude that $F^{03}|_{\rm R_+}= 
F^{03}|_{\rm R_-}$.  But since  $F^{03}=f(r)e^{\imath(m\phi-\omega t)}$, it
is obvious that  $f$ must be continuous at $r=R$.   By similarly integrating
Eq.(\ref{source}d) and invoking the boundedness of $H^{23}, H^{03}$ and $E_z$
we find $H^{31}|_{\rm R_+}=  H^{31}|_{\rm R_-}$.  Then from
Eqs.~(\ref{algebraic}b) and (\ref{true}a) it follows that $f'|_{\rm
R_+}=(f'/\mu)|_{\rm R_-}$.   One checks that with these matching conditions
$\overline S_r$ in Eq.~(\ref{energyflux}) is continuous at
$r=R$.

With the expressions for $f_{\rm in}$ and $f_{\rm in}$ written out earlier, 
the matching conditions are
\begin{mathletters}
\label{match}
\begin{equation}
\tau J_m(\kappa R) = \rho H^{(1)}_m(\omega R) +  H^{(2)}_m(\omega R)
\end{equation}
\begin{equation}
(\tau\kappa/\mu) J_m{}'(\kappa R) = \rho \omega H^{(1)}_m{}'(\omega R) + 
\omega H^{(2)}_m{}'(\omega R)
\end{equation}
\end{mathletters}
where $'$ here means derivative with respective to the argument.  Solving
these simultaneously for $\rho$ and rearranging the result with help of the
identity  $H^{(1)}{}_m = H^{(2)}_m{}^*$ gives
\begin{mathletters}
\label{rho}
\begin{equation}
\rho = -{1-\mu\chi_m(x) \eta_m(y)\over 1-\mu\chi_m(x)^* \eta_m(y)}\cdot{
H^{(2)}_m(x)\over  H^{(2)}_m(x)^*}
\end{equation}
\begin{equation}
\chi_m(x) \equiv x  H^{(2)}{}_m'(x)/ H^{(2)}{}_m(x);\quad \eta_m(y)
\equiv J_m(y)/[y J_m'(y)]
\end{equation} 
\end{mathletters}
with $x\equiv \omega R$ and  $y\equiv \kappa R$.

When there is no dissipation, $\epsilon$ and $\mu$ are real while $\sigma=0$,
and so $y$ is real.  It follows that numerator and denominator of
Eq.~(\ref{rho}a) are complex conjugates so that $|\rho|=1$.  This is in
harmony with the arguments of Sec.~\ref{rotational} that superradiance goes
hand in hand with dissipation.    

Let us now define the dimensionless parameters $v\equiv \Omega R$
(peripheral velocity of the cylinder in units of $c$) and $\xi \equiv
4\pi\mu\sigma R$ in terms of which
\begin{equation}
y^2 = x^2+(1-\epsilon\mu) (x-mv)^2 \gamma^2 + \imath\, (x-mv)\xi\gamma
\label{y}
\end{equation}
The $y$ shall be the square root which is positive in the limit
$\sigma\rightarrow 0$.   A useful approximation for the gain coefficient $-a_m
\equiv |\rho|^2-1$ [this is the same as the coefficient $a_m(\omega)$
appearing in Sec.~\ref{rotational}] can be obtained from Eq.~(\ref{rho}) by
passing to the nonrelativistic limit $v\ll 1$, $\gamma\approx 1$ which, for
$m$ not too large, implies  $x\ll 1$.

First the recursion relation
\cite{Abramowitz} $xH_m^{(2)}{}'=xH_{m-1}^{(2)}-mH_m{}^{(2)}$ allows us to
write
\begin{equation}
\chi_m(x) = -m + {xH_{m-1}^{(2)}(x)\over H_{m}^{(2)}(x)}
\label{newchi}
\end{equation}
For $x\ll 1$ the leading terms of the real and imaginary parts of the Hankel
function are \cite{Abramowitz}
\begin{mathletters}
\label{Hankel}
\begin{equation}
H_0^{(2)} \approx 1 - {2\imath\over \pi}(\ln{x\over 2} +\gamma_E)
\end{equation}
\begin{equation}
H_m^{(2)}(x) \approx {x^m\over m! 2^m} + \imath {2^m\over \pi x^m};\quad m
\geq 1
\end{equation}
\end{mathletters}
where $\gamma_E\approx 0.577216$ is the Euler--Mascheroni constant. 
Substituting in Eq.~(\ref{newchi}) we have to leading real and imaginary
orders in $x$
\begin{equation}
\chi_m(x) \approx - m - \delta_m^1\left({1\over 2} + \ln {x\over 2} +
\gamma_E\right)x^2 + {x^2\over 2} -{\imath\pi x^{2m}\over (m-1)!
2^{2m-1}} + \cdots
\label{chicalc}
\end{equation}

We now substitute from Eq.~(\ref{chicalc}) into Eq.~(\ref{rho}a) and
recall that the ratio of $H_m^{(2)}$ to its complex conjugate has unit
modulus.  Factoring out $1+ \mu m \eta_m(y)$ from numerator and denominator,
we find in each the function $h_m(y)\equiv \mu\eta_m(y)[1+\mu m
\eta_m(y)]^{-1}$ multiplied in one by a small complex expression and in the
other by the conjugate of this expression.  As a result to leading
$\left[{\cal O}(x^2)\right]$ order, $h_m$ appears in $\rho$ multiplied only by
an imaginary factor, so that only the imaginary part of $h_m$ remains in
$|\rho|^2$.  Retaining only dominant terms leads to
\begin{equation}
a_m  \approx {8\pi (x/2)^{2m}\over (m-1)!}\,\Im\,
h_m(y) =  {8\pi (x/2)^{2m}\over (m-1)!}\,\Im\,
{\mu J_m(y)\over (\mu-1)m J_m(y)+y J_{m-1}(y)}
\label{a1}
\end{equation}
where the last form follows from the recursion relation \cite{Abramowitz}
$yJ_m'=yJ_{m-1}-mJ_m$.  Since $-a_m$ is proportional to the small factor
$x^{2m}$, superradiance is mostly confined to the $m=1$ mode (unless the
ingoing wave only has $m>1$).

We went through the derivation of Eq.~(\ref{a1}) with possibly
complex $\epsilon$ and $\mu$ as a matter of principle, and because it will be
required for the discussion in the Appendix.  But in practice
little need can arise to consider complex $\epsilon$ or $\mu$.   For low
frequencies both these quantities are real with $\epsilon$ becoming complex in
real materials only at frequencies $ > 10^{11}$ Hz (in ferromagnets $\mu$ can
become dispersive at somewhat lower frequencies) \cite{LLECM}.  Recall that
the appropriate argument of $\epsilon$ or $\mu$ in our discussion is $\omega -
m\Omega$ which must be negative.  But a macroscopic cylinder rotating
nonrelativistically will do so below $\Omega=10^{10}$ Hz.  And as mentioned,
$m$ cannot be large without superradiance being suppressed. Thus in the
laboratory we cannot arrange for $\omega - m\Omega$ to be negative and
sufficiently large in magnitude to access the complex range of $\epsilon$ or
$\mu$. Henceforth we consider only real $\epsilon$ and $\mu$.

As mentioned, for nonrelativistic rotation $v\ll 1$ and $x\ll 1$ and thus
$|x-mv|\ll 1$.  The low conductivity regime may be defined by the additional
condition
\begin{equation}
 |x-mv|\xi \ll 1
\label{cond}
\end{equation}
When all these are valid, the argument $y$ of the Bessel functions is a small
{\it complex\/} number, and we can expand
\begin{equation}
J_m(y)= {y^m\over 2^m\, m!} \left[1-{y^2\over 4(m+1)}+ \cdots\right]
\label{bessel}
\end{equation}
Substituting this, Eq.~(\ref{y}) and the definitions of $\xi, x$ and $v$ in
Eq.~(\ref{a1}) and reinstating $c$ gives to leading order
\begin{equation}
a_m \approx {16\pi^2\mu^2 (\omega R/2c)^{2m}(\omega-m\Omega)\sigma
R^2/c^2 \over m(m+1)!
(\mu+1)^2\,}\cdot
\label{am}
\end{equation}
which shows clearly that for $(\omega-m\Omega)<0$ there is superradiance
($a_m<0$).  The formula supports Zel'dovich's assertion that for low
conductivity the gain coefficient is proportional to $\sigma
R^2(m\Omega-\omega)(\omega R)^{2m}$.  Our result gives the proportionality
constant and shows that $-a_m$ is independent of $\epsilon$.  Numerical work
shows that Eq.~(\ref{am}) remains accurate to within $1\%$ up to $ |x-mv|\xi
\approx 1$.

For $ |x-mv|\xi > 1$ we return to Eq.~(\ref{a1}).  Because the gain 
coefficient falls off with growing $m$, we discuss here only the results for
$m=1$.  Clearly the terms $x^2$ and $(x-mv)^2$ in $y^2$ are negligible because
$x<v$ and $v\ll 1$.  (We presume that $\epsilon\mu$ is not too large, which
is reasonable because for a good conductor $\epsilon$ and $\mu$ are formally
unity).  Hence the argument $y$ in Eq.~(\ref{a1}) reduces to
$[(x-v)\xi]^{1/2}$.  The imaginary part is best evaluated numerically.  As a
function of $(x-v)\xi$ it sports single maximum of height
$0.1887$ located  at $(x-v)\xi\approx -6.325$.   From these last numbers and
Eq.~(\ref{a1}) we infer the maximal gain coefficient for given $\omega$:
\begin{equation}
-(a_1)_{\rm max} = 1.185(\omega R/c)^2\quad {\rm at }\quad \Omega = \omega
+0.503 c^2/ \sigma R^2
\label{optimal}
\end{equation}
For a copper cylinder with $R=10$ cm, the minimum $\Omega$ required for the
peak to be present is 0.06 s$^{-1}$; this is also the offset between
$\omega$ and the $\Omega$ giving maximum superradiance.

\subsection{Rotational Superradiance Devices} \label{devices}

From Eqs.~(\ref{am})--(\ref{optimal}) it is clear that for
superradiance of a nonrelativistically rotating cylinder the gain coefficient
$-a_m$ is extremely small (basically $\omega R/c$ is very small).  This would
seem to imply that superradiance cannot be observed in the laboratory.  But in
fact this is not the case for two reasons.  First by surrounding the cylinder
with a jacket made of material where the speed of light is rather small, one
achieves a more favorable ratio of cylinder radius to wavelength with a
consequent  improvement in $-a_m$.   Second, a suitable device can cycle the
amplified radiation any number of times to compound the gain
coefficient.  This raises the possibility of practical devices for
amplification of signals at the expense of mechanical energy.

To explain the reason for the first improvement in the simplest terms we
consider the jacket material to have $\mu=1$ but very large and {\it
real\/} permittivity $\epsilon_{\rm j}$.  Eqs.~(\ref{algebraic})--(\ref{source})
can obviously be used outside the cylinder if we put everywhere
$\sigma=\varrho=\Omega=0$.  The arguments of Sec.~\ref{modes} characterizing
the AE and AM modes can be repeated with like conclusions.  For AE modes
we need to replace the radial equation (\ref{radialoutside}) outside the
cylinder by ({\it c.f.\/} Eq.~\ref{radial2})
\begin{equation}
r^2 f'' + rf' -[m^2-\epsilon_{\rm j}\,\omega^2 r^2]f=0; \quad r> R
\label{radialoutside2}
\end{equation}
Therefore, the argument of the Hankel functions in Sec.~\ref{amplification} is
now $\surd\epsilon_{\rm j}\, \omega r$ rather than $\omega r$.  And the
Hankel and $\chi_m$ functions in Eqs.~(\ref{match})-(\ref{rho}) now take
argument $\surd\epsilon_{\rm j}\,x$.  There is no change in the
matching conditions $f|_{\rm R_+}=f|_{\rm R_-}$ and 
$f'|_{\rm R_+}=(f'/\mu)|_{\rm R_-}$ since $\mu$ has not been changed.  In
Eq.~(\ref{match}b) a factor $\surd\epsilon_{\rm j}$ appears alongside $\mu$;
it comes from the arguments of the differentiated Hankel functions.  No
change occurs in $y$, the argument of the Bessel functions, which is composed
exclusively of quantities describing the cylinder.

Let us assume that even though $\epsilon_{\rm j}$ is large,
$\surd\epsilon_{\rm j}\, x\ll 1$ (remember we are in the superradiant regime
so $\Omega R\ll 1$). The assumption means that the rotational velocity is still
well below the speed of light in the jacket.  Then we can expand the Hankel
functions for small argument as before and arrive back at formulas (\ref{am})
and (\ref{optimal}) with the replacements $\mu\rightarrow
\mu\surd\epsilon_{\rm j}$ and $x\rightarrow
\surd\epsilon_{\rm j} \,x$.  Since $\surd\epsilon_{\rm j}$ is assumed
large, the $\mu$ dependent factor in Eq.~(\ref{am}) is here replaced by unity
so that
\begin{mathletters}
\label{am2}
\begin{equation}
a_m \approx {16\pi^2(\epsilon_{\rm j})^m (\omega
R/2c)^{2m}(\omega-m\Omega)\sigma R^2/c^2 \over m(m+1)!}\cdot
\end{equation}
\begin{equation}
-(a_1)_{\rm max} = 1.185\,\epsilon_{\rm j}\,(\omega R/c)^2\quad {\rm at }\quad
\Omega = \omega + 0.503 c^2/ \sigma R^2
\end{equation}
\end{mathletters}
In Eq.~(\ref{am2}a) the factor $\omega-m\Omega$ is unchanged because it
stems from $y$. Thus a jacket of high $\epsilon_{\rm j}$ material provides,
for $m=1$, a gain larger by a factor $\epsilon_{\rm j}$ over the
vacuum value.

The second ingredient of the superradiant device is cycling through
reflection.   Suppose the rotating cylinder and its high--$\epsilon$ jacket
are placed inside a concentric cylindrical reflecting cavity of radius
$R_c>R$ (this is similar to Press and Teukolsky's idea for the ``black hole
bomb'' \cite{PressTeuk2}).  Introduce in the intervening material an
electromagnetic wave with low
$m$ components.  One simple way to do this is to apply across the ends of the
cylinder along one edge a voltage  varying sinusoidally with frequency
$\omega$; this will produce preferentially low $m$ waves with their electric
field parallel to the cylinder's axis (hence AE modes).  Each such wave which
satisfies the  Zel'dovich--Misner condition gains in power as per
Eq.~(\ref{am2}) as it interacts with the cylinder.   Propagating out, the
amplified wave is reflected back by the cavity for a second round of
amplification, and so on.  If the cavity is a perfect reflector, and the
material between cylinder and cavity is perfectly transparent, there will be a
net gain in power which increases linearly with the number of bounces. 
But if the cavity absorbs (or leaks radiation outward), the consequent loss in
power may quench the process.  However, absorption in the cavity may be
turned to our advantage by making the cavity rotate in the same sense as the
cylinder with sufficiently large angular frequency so as to cause it also to
superradiate for the modes in question.   If the cavity walls are thick
enough to prevent leakage, then  each of the waves mentioned  will always
gain power in each round trip, and the overall gain is limited only by the
time one allows the process go on.

When estimating the efficiency of such devices, the principal
question is how big can $-a_m$ be.  For an isolated cylinder, and $m=1$ AE
modes, Eq.~(\ref{am2}b) gives for optimal parameters that
$-(a_1)_{\rm max} \approx 1.2\, \epsilon_{\rm j}\,(\omega R/c)^2$.  For
the cylinder--cavity device, this optimum gain is acquired over the
back-and-forth light travel time $2\,\surd\epsilon_{\rm j}\,(R_c-R)/c$; one
must still add to it the gain due to the cavity.  As mentioned, for a
cylinder made of good conductor, the peak superradiance occurs at
$\Omega\approx\omega$.  Hence the e--folding time of the cylinder--cavity
device is $T_e <1.67 c(R_c-R)/(\surd\epsilon_{\rm j}\,\Omega^2 R^2)$.  With
$R_c=2R=20$ cm and $\Omega=2\pi\times 10^2$ s$^{-1}$, $T_e\approx
(4/\surd\epsilon_{\rm j})\,$hour, so that the effect can become dramatic for
large $\epsilon_{\rm j}$.  Many materials made of polar molecules have big
$\epsilon$ at low frequencies, {\it e.g.\/} $\epsilon(0) \approx 80$ for
water ice while  $\epsilon(0) \approx 300$ for lead telluride
\cite{Kittel}. And ferroelectrics just above the Curie point have virtually
unbounded $\epsilon(0)$ \cite{LLECM}.

A variation on the above is to have a coaxial cable (with no filling) rotating
about its axis.  Wave modes which not only have angular variation, but also
vary along the axis (the $k\neq 0$ case studied in Sec.~\ref{equations})
will travel along the cable while bouncing between inner and outer
boundaries.  So long as  the Zel'dovich--Misner condition is satisfied for
such a mode, it will be amplified - rather than damped - as it travels along
the cable. This might prove useful in protecting signals from degradation.  We
should stress that similar amplification will take place whatever the nature
of the wave, sound waves being another useful candidate.  

\acknowledgments

JDB thanks Mordehai Milgrom for many discussions, and George Blumenthal for
pointing out a possible source of confusion.  MS thanks the Racah Institute of
Physics for hospitality, and the FAPESP for support.  This work is
supported by a grant from the Israel Science Foundation.

\appendix
\section{Superradiance in Axial Magnetic Modes}
\label{long_magnetic}

For completeness we now work out the gain coefficient for the AM
modes with $k=0$.  We set $\epsilon=\mu=1$ inside the cylinder to simplify the
equations.  Thus $H^{\alpha\beta}=F^{\alpha\beta}$ everywhere.  By the
definition of the modes we have $F^{03}=F^{31}=F^{23}=0$. 

We combine Eqs.(\ref{source}a,c) judiciously to cause the charge density terms
(wherever nonvanishing) to cancel:
\begin{mathletters}
\label{combine}
\begin{equation}
\partial(F^{12}+\Omega F^{01})r/\partial r + \zeta^{-1} r F^{02} =0
\end{equation}
\begin{equation}
\zeta^{-1} \equiv  4\pi\sigma \gamma^{-1}\Theta(R-r) -\imath(\omega-m\Omega)
\end{equation}
\end{mathletters}
Here $\Theta$ denotes the Heavyside step function.  The function $g(r)\equiv
(F^{12}+\Omega F^{01})r e^{-\imath(m\phi-\omega t)}$ shall here play a role
analogous to $f(r)$ in Sec.~\ref{amplification}.  In terms of it
Eq.~(\ref{combine}a) gives 
\begin{equation}
 F^{02} = - \zeta r^{-1} g' e^{\imath(m\phi-\omega t)}
\label{derivativeg}
\end{equation}
Now for $r<R$ we eliminate $E_r$ between Eqs.(\ref{algebraic}c)
and (\ref{source}b) to obtain $F^{01}/F^{12}$ so that we may express $g$ in
terms of $F^{12}$ alone.  It follows that
\begin{mathletters}
\label{F12}
\begin{equation}
F^{12} =  (4\pi\sigma\gamma -\imath\omega) \zeta r^{-1} g
e^{\imath(m\phi-\omega t)}
\end{equation}
\begin{equation}
F^{01} = - (4\pi\sigma\gamma\Omega r^2 -\imath m) \zeta r^{-1} g
e^{\imath(m\phi-\omega t)}
\end{equation} 
\end{mathletters}
For $r>R$ we use solely Eq.~(\ref{source}b) to determine $F^{01}/F^{12}$; the
result is again Eqs.~(\ref{F12}a,b) with $\sigma\rightarrow 0$. Hence all
nonvanishing field components can be recovered from $g$.  

Substituting all these results in Eq.~(\ref{newhomogeneous}a) we get the
radial equation for the AM modes:
\begin{equation}
\left( \zeta r g' \right)' -\zeta \left[m^2-\omega^2 r^2
-\imath 4\pi\gamma\sigma  (\omega-m\Omega) r^2\Theta(R-r) \right] r^{-1} g =0
\label{equationg}
\end{equation}
Because $\epsilon=\mu=1$ here, this equation is
quite similar to that for $f$, Eq.~(\ref{inside2}); in fact the only
difference between them is a term involving $d\zeta/dr$.  This last will
vanish in the nonrelativistic limit where $\zeta$ becomes constant (except at
$r=R$), and in that limit the equations are identical both inside and outside
the cylinder.  Indeed  
\begin{mathletters}
\label{gequation}
\begin{equation}
r^2 g'' + rg' -[m^2-\omega^2 r^2]g=0; \quad r> R
\end{equation}
\begin{equation}
r^2 g'' + rg' -[m^2-\tilde\kappa^2 r^2]g=0; \quad r< R
\end{equation}
\begin{equation}
\tilde\kappa^2 \equiv \omega^2 +\imath 4\pi\sigma (\omega-m\Omega)
\end{equation}
\end{mathletters}

By analogy with Sec.~\ref{amplification} the solution outside the cylinder
is $g|_{R_+} =  H_m^{(2)}(\omega r)+\rho H_m^{(1)}(\omega r)$ while that
inside is $g|_{R_-} =\tau J_m(\tilde\kappa r)$.  To find the matching
conditions at $r=R$ we note that the condition of continuity of tangential
electric fields requires that $F^{02}|_{R_+}=F^{02}|_{R_-}$. 
By Eq.~(\ref{derivativeg}) this means $(\zeta g')|_{R_+}=(\zeta g')|_{R_-}$.
Further, by integrating Eq.~(\ref{combine}) over a small radial interval
spanning $r=R$ and realizing that all quantities are bounded, we see that
$g|_{R_+}=g|_{R_-}$.  These matching conditions parallel those for $f$
when one replaces $\mu|_{R_-}\rightarrow \zeta^{-1}|_{R_-}$ and
$\mu|_{R_+}\rightarrow \zeta^{-1}|_{R_+}$.  Recalling
Eq.~(\ref{combine}b) we see that Eqs.~(\ref{match}), (\ref{rho})  and
(\ref{a1}) are applicable here with the replacements
$\kappa\rightarrow\tilde\kappa$,
$y\rightarrow \tilde y$ and 
$\mu\rightarrow \tilde\mu$, where
\begin{mathletters}
\label{bars}
\begin{equation}
\tilde y^2 \equiv x^2+\imath \xi(x-mv)
\end{equation}
\begin{equation}
\tilde\mu=\imath(\omega-m\Omega)^{-1}\,\zeta^{-1}|_{R_-}=1+\imath\xi(x-mv)^{-1}
\end{equation}
\end{mathletters}

We now obtain a formula analogous to (\ref{am}) valid for $x\ll 1$ and when
the small conductivity  condition~(\ref{cond}) holds.  We substitute the
expansion (\ref{bessel}) into Eq.~(\ref{a1}) and retain terms to O$(\tilde
y^2)$. The isolation of the imaginary part is easier if the denominator is
put in real form.  Neglecting terms in the numerator of higher order in $x$
and $x-mv$, and reverting to dimensional quantities we get
\begin{equation}
\label{amnew}
a_m \approx {8\pi^2 (\omega R/2c)^{2m}
\over m!} { (\omega-m\Omega)\sigma\over (\omega-m\Omega)^2 + 4\pi^2\sigma^2}
\end{equation}

This formula again shows that superradiance occurs only for
$\omega-m\Omega<0$, and is in harmony with the expansion
(\ref{newcondition}).  It supports the insight mentioned in
Sec.~\ref{amplification} that superradiance is significant only for $m=1$. 
It corrects  Zel'dovich's approximate formula for the AM modes, $a_m\propto
(\omega-m\Omega) \sigma [(\omega-m\Omega)^2 +16\pi^2\sigma^2)]^{-1}$ and
supplies the normalization. We note that for fixed $\omega R$, $a_1$ has the
peak
\begin{equation}
-(a_1)_{\rm max} = 1.571(\omega R/c)^2\quad {\rm at }\quad \Omega = \omega
+2\pi\sigma
\label{optimal2}
\end{equation}
This peak gain is similar to that for AE modes.  But unless the cylinder's
conductivity is small, the $\Omega$ required to reach the peak gain will not
be a practical one.  For example, for copper $\sigma \approx 10^{17} s^{-1}$.
Put another way, for given $v$ the peak is accessible only if $\xi<2v$.  For
larger $\xi$ we must resort to numerical evaluation of the imaginary part in
Eq.~(\ref{a1}) with the substitutions (\ref{bars}); it certifies that the
peak gain (\ref{optimal2}) is not even approached.  In closing we should note
that for small $\xi$ faster rotation is necessary to reach the peak gain for
AE modes than for AM modes.

\end{document}